\documentclass[12pt]{article}
\usepackage{authblk}
\usepackage{geometry}
\usepackage{amsmath}               
\usepackage{amssymb}
\usepackage{amsthm}
\usepackage{mathrsfs}
\usepackage{customcommands}
\usepackage{bm}
\usepackage{Nettastyle}
\usepackage{dsfont}
\usepackage{comment}
\usepackage[utf8]{inputenc}
\usepackage{calc}
\usepackage{accents}
\usepackage{amsmath}
\usepackage{amssymb}
\usepackage{amsfonts}

\newtheorem{assump}{Assumption}
\usepackage{braket}
\usepackage[normalem]{ulem}
\usepackage{color}
\usepackage{bbold}
\usepackage{ulem}
\usepackage{mathtools}
\usepackage{tensor} 
\usepackage{tikz}
\usetikzlibrary{arrows, positioning, quotes, intersections}
\usepackage{graphicx}

\preprint{MIT-CTP/5740}

\newcommand{\assign}[1]{\textcolor{green}{Needs Volunteer}}

\DeclareMathOperator{\tr}{tr}

\title{Spoofing Entanglement in Holography}

\author[1]{Netta Engelhardt,}
\author[1]{{\AA}smund Folkestad,}
\author[1]{Adam Levine,}
\author[1,2]{Evita Verheijden}
\author[1,3]{and Lisa Yang}

\affiliation[1]{Center for Theoretical Physics, Massachusetts Institute of Technology, \\Cambridge, MA 02139, USA}
\affiliation[2]{Black Hole Initiative at Harvard University, \\ 20 Garden Street, Cambridge, MA 02138, USA}
\affiliation[3]{Computer Science and Artificial Intelligence Laboratory,\\
Massachusetts Institute of Technology, Cambridge, MA 02139, USA}
\emailAdd{engeln@mit.edu}
\emailAdd{afolkest@mit.edu}
\emailAdd{arlevine@mit.edu}
\emailAdd{evitamhv@mit.edu}
\emailAdd{lisayang@mit.edu}

\abstract{
A defining property of Hawking radiation is that states with very low entanglement masquerade as highly mixed states; this property is captured by a quantum computational phenomenon known as spoofing entanglement. 
Motivated by the potential implications for black hole information and the emergence of spacetime connectivity, as well as possible applications of spoofing entanglement, we investigate the geometrization of two types of entanglement spoofers in AdS/CFT: so-called EFI pairs and pseudoentangled state ensembles. We show that (a strengthened version of) EFI pairs with a semiclassical bulk dual have a Python's Lunch; the maximally mixed state over the pseudoentangled state ensemble likewise features a Python's Lunch. Since a Python's Lunch must lie behind an event horizon, we find that black holes are the exclusive  gravitational source of entanglement spoofing in the semiclassical limit. Finally, we use an extant construction of holographic pseudorandom states to yield a candidate example of a pseudoentangled state ensemble with a semiclassical bulk dual.} 

\begin{document}

%begin macros
%%%% shortcut macros
\renewcommand{\[}{\left[}
\renewcommand{\]}{\right]}
\renewcommand{\(}{\left(}
\renewcommand{\)}{\right)}
\renewcommand{\gets}{\leftarrow}

%math/CS macros
\newcommand{\Nat}{\mathbb{N}} %natural numbers 
\newcommand{\poly}{\mathrm{poly}}
\newcommand{\Hn}{n} %number of qubits in CS
\newcommand{\Hd}{d} %dimension of Hilbert space in CS 
\newcommand{\secp}{\kappa} %security parameter in cryptography 
\newcommand{\avg}{\mathrm{avg}}
\newcommand{\avgover}[1]{\underset{#1}{\mathrm{avg}}} 
\newcommand{\const}{\alpha} %constant in the fidelity bound 
\newcommand{\POVM}{E} %POVM that \Alg cannot reconstruct
\newcommand{\operator}{\mathcal{O}} %operator \Alg cannot reconstruct
\newcommand{\negl}{\mathrm{negl}}

\newcommand{\Hil}{\mathcal{H}} %Hilbert space 
\newcommand{\key}{k} %key of PRU/PRS
\newcommand{\PRU}{U} %PRU 
\newcommand{\PRUens}{\mathcal{U}} %PRU ensemble
\newcommand{\PRS}{\psi} %PRS density operator 
\newcommand{\PRSens}{\Psi} %PRS ensemble
\newcommand{\ketPRS}{\ket{\psi}} %PRS ket
\newcommand{\Alg}{\mathcal{A}} %learning/reconstruction algorithm 
\newcommand{\avgf}[1]{{\langle F \rangle^{#1}}} %average fidelity for algorithm

\newcommand{\HN}{N} %CFT parameter, internal degrees of freedom 
\newcommand{\op}{O} %operator

\maketitle
\section{Introduction} 

A central aspect of the black hole information paradox is that the Hawking radiation after the Page time appears to be more entangled than it has any right to be \cite{Haw76,Pag93b}. After the Page time, the state of the radiation $\Psi_{\rm rad}$ is computationally indistinguishable from a highly mixed (thermal) state \cite{Pag93a, HarHay13}: the unitary dynamics of black hole evaporation generate entanglement-spoofing states from ordinary states. This is an example of a more general phenomenon known as \textit{spoofing entanglement}, in which low-entanglement states masquerade as highly entangled states (and vice versa).  What are the necessary conditions required by unitary operators to generate such states? How typical is this phenomenon? Are black holes the only systems whose time evolution generates states that spoof an ${\cal O}(1/G_{N})$ amount of entanglement, or is there an analogue of this aspect of the information problem in other gravitational states? What does spoofing entanglement imply for spacetime emergence? In holography, entanglement is expected to build spacetime connectivity \cite{Van09, Van10, MalSus13, EngLiu23} when the bulk  entanglement term is subleading \cite{EngFol22}; in this regime, the absence of leading order entanglement forbids connectivity between two asymptotic boundaries  (since the QES~\cite{RyuTak06, HubRan07, FauLew13, EngWal14} of each will be empty~\cite{EngWal14}). 
If actual entanglement implies spacetime connectivity, what does spoofing entanglement imply?

Here we initiate an investigation into these questions by identifying the bulk geometrization of boundary entanglement-spoofing states at leading order in $N$. 
As one of our conclusions, we find that computationally indistinguishable states have identical outer wedges~\cite{EngWal17b}. Furthermore, we show that one of the states must have a  geometric feature known as a Python's Lunch~\cite{BroGha19}, the bulk avatar of exponential reconstruction complexity. It is thus exponentially complex to assess potential spacetime connectivity: the Python's Lunch spoofs spacetime connectivity.

Spoofing entanglement generically entails three properties: first, there are low entanglement states and high entanglement states; second, these states are computationally indistinguishable from each other (in a sense to be made precise below); and third, both (sets of) states are simple to prepare --- a property we refer to as ``efficiently preparable''. We will be interested in two types of entanglement spoofers: EFI pairs \cite{BraCan22} and pseudoentangled state (PES) ensembles \cite{AarBou22}, which we now briefly introduce.\footnote{Note that other types of entanglement spoofers exist; see, e.g.,~\cite{ArnBra23} and~\cite{BouFef23}.}

EFI pairs are pairs of efficiently preparable mixed states that are far in trace distance, but computationally indistinguishable given a \textit{single} copy of each state.\footnote{EFI stands for Efficiently preparable, statistically Far, and computationally Indistinguishable. We give a review in Sec.~\ref{sec:review}; a technical definition can be found in \cite{BraCan22}.} PES ensembles are pairs of efficiently preparable \textit{ensembles} of quantum states with respectively low and high entanglement across a bipartition.\footnote{In the second version of the work of \cite{AarBou22}, the definition of pseudoentangled states was expanded to require that the states are pseudoentangled across almost all partitions of the system. In this work, we follow the first version of \cite{AarBou22}, in which the states are pseudoentangled across a single, fixed cut, because we think it is the only reasonable notion in holography. We will comment further on this in the discussion section.} Furthermore, the ensembles are ``indistinguishable'': to any computationally bounded observer the states appear to be equivalent, even if the observer is given access to polynomially many copies of the states. 

We emphasize that while EFI pairs are a single pair of states, as opposed to two ensembles of states for PES ensembles, neither notion of entanglement spoofing implies the other. On the one hand, EFIs are a stronger notion as there is no randomness that would come from sampling from ensembles facilitating their indistinguishability. On the other hand, EFI pairs are also weaker than PES ensembles: they are only guaranteed to be indistinguishable given a single copy of the state, whereas PES ensembles are indistinguishable given polynomially many copies of a state drawn from the ensemble. For this reason, PES ensembles obey a lower bound on the entropy of states in the less-entangled ensemble; for EFI pairs, there is a lower bound instead on the rank of the mixed states. In particular, EFI pairs are actually a more general concept: they do not necessarily have to consist of a pair of low and high entanglement states, but they do include any such pair of entanglement spoofers.  

We consider ``geometrically entangled states'': states in which  the entropy of bulk quantum fields is $\mathcal{O}(1)$, so that any $\mathcal{O}(1/G_N)$ amount of entanglement is sourced by an area term. Our primary workhorse is the strong Python's Lunch proposal~\cite{EngPen21b, EngPen23}, which identifies bulk operators whose reconstruction requires exponential complexity in $1/G_N$. We now briefly review this proposal. Consider some code subspace ${\cal H}_{\rm code}$. Ref.~\cite{BroGha19} argued that reconstruction complexity of bulk operators behind the outermost QES in any state in ${\cal H}_{\rm code}$ scales exponentially in  $1/G_N$ (in the case of a generic classical Python).  More generally, reconstruction is exponentially complex behind the outermost QES of the maximally mixed state on ${\cal H}_{\rm code}$, even if other states do \textit{not} have any nonempty QES (see \eqref{eq:Cpython} for the precise exponent). 

\begin{figure}
        \centering
        \includegraphics[scale=0.7]{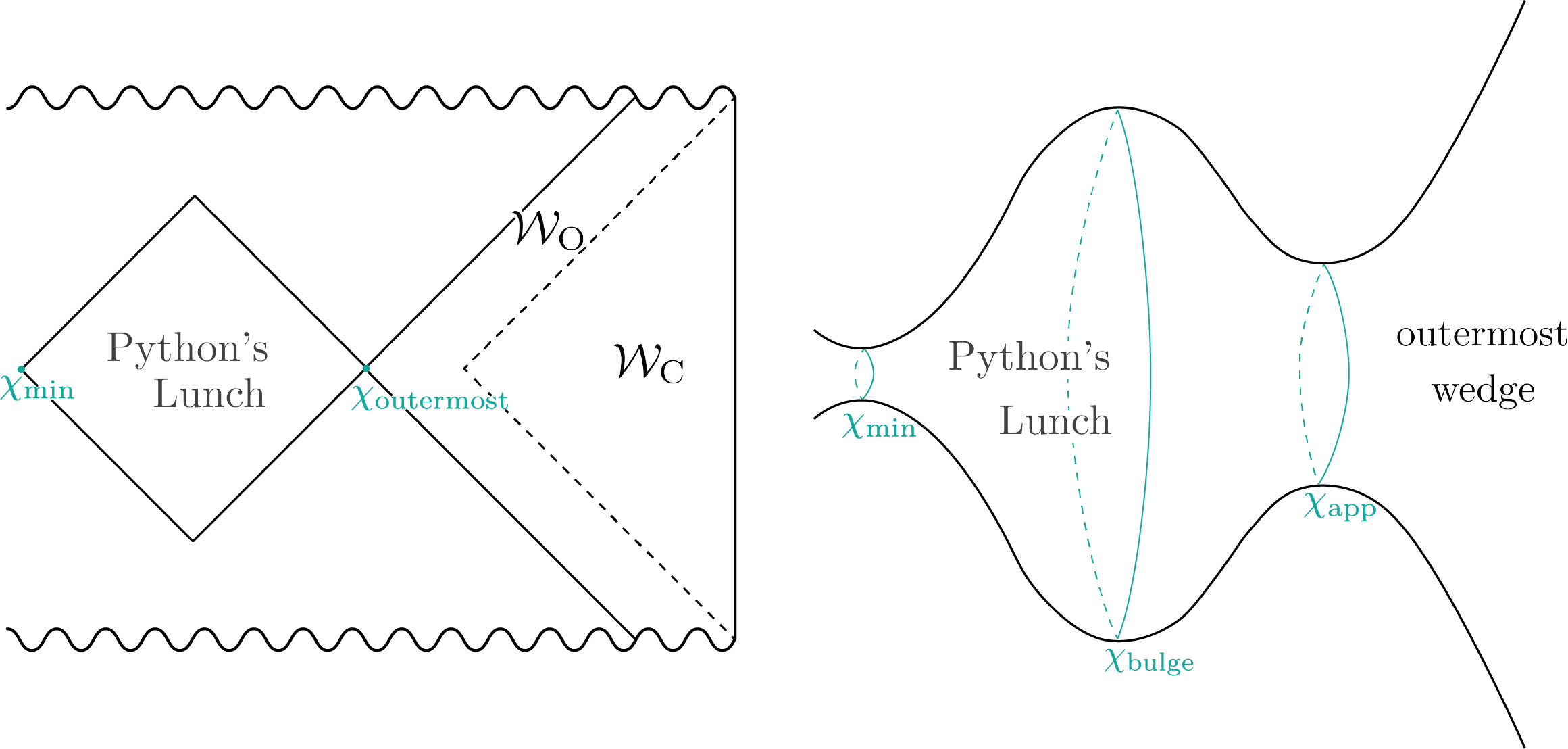}
        \caption{A spacetime with a Python’s Lunch. In the Penrose diagram on the left, the Lunch, causal wedge ($\mathcal{W}_C$) and the wedge of the outermost QES ($\mathcal{W}_O$) are indicated. The left and right green dots are the minimal and outermost QES, respectively. The right panel depicts a timeslice.}
        \label{fig:Python}
    \end{figure}

Concretely, if $\chi_{\rm app}$ (the so-called ``appetizer") is the outermost QES in the entanglement wedge of some complete asymptotic boundary, and $\chi_{\rm app}$ is \textit{not} identical to the minimal QES $\chi_{\rm min}$ defining the entanglement wedge, then it is possible to prove that a third QES $\chi_{\rm bulge}$ lives behind $\chi_{\rm app}$~\cite{BroGha19}. The strong Python's Lunch conjecture states that operators localized between the asymptotic boundary and $\chi_{\rm app}$ admit a polynomially (or subexponentially, the distinction is immaterial for us) complex reconstruction. This region is known as the ``simple wedge'' (also called the outer wedge)~\cite{EngWal17b, EngWal18}. Operators with support behind the appetizer are exponentially complex to reconstruct, in $\log |{\cal H}_{\rm code}|$; we shall refer to this region as the Lunch. In terms of the generalized entropies of the QESs, the reconstruction complexity of operators inside the Lunch is 
\begin{equation}\label{eq:Cpython}
    C\sim {\rm exp}\left [ \frac{1}{2}\left ( S_{\rm gen}[\chi_{\rm bulge}]-S_{\rm gen}[\chi_{\rm app}]\right) \right].
\end{equation}
See Fig.~\ref{fig:Python} for an illustration of this geometry.\footnote{The reader may note at this point that for purposes of spoofing entanglement, we are interested not in complexity of reconstruction but complexity of \textit{distinguishing}, which at least \textit{prima facie} appears to be a less difficult task. However, as argued in footnote 32 of~\cite{AkeEng22}, distinguishing is not much simpler than reconstruction. Indeed, in the concrete tensor network models of~\cite{AkeEng22}, exponential complexity of distinguishing was found whenever there was a Python's Lunch in the network. We conjecture that \eqref{eq:Cpython} also measures the complexity of distinguishing.}
Generically, $S_{\rm gen}[\chi_{\rm bulge}]-S_{\rm gen}[\chi_{\rm app}]\sim {\cal O}(1/G_N)$, so any operators with support behind $\chi _{\rm app}$ will generically have reconstruction complexity that scales exponentially in $1/G_N$. 

We argue that for geometrically entangled holographic states to spoof entanglement (as measured by $S_{\rm vN}$) at leading order in $N$,\footnote{Note that there is no other natural parameter in the CFT (except possibly $\lambda$) that one could use to quantify the amount of pseudoentanglement.} the bulk must have a Python's Lunch. In particular, we consider a slightly stronger notion of EFIs than originally defined in \cite{BraCan22}: our ``large-$N$ EFIs'' are further apart in trace distance than ordinary EFIs and are computationally indistinguishable with access to any $\mathcal{O}(1)$ number of copies of the state. For reasons elucidated below, we find these to be more appropriate to the holographic setting. For these stronger EFI pairs we show that at least one of the two states in the pair must have a Python's Lunch in the bulk dual. In fact, as noted above, we find that multi-copy computational indistinguishability generically implies identical outer wedges. Our second example of entanglement-spoofing states in holography are PES ensembles, and we find that given a pair of PES ensembles, at least one of the ensembles must have a Python's Lunch in the maximally mixed state (when they have a semiclassical bulk dual).\footnote{Note that in the case of EFI pairs, it is possible that the maximally mixed state on the code subspace \textit{also} has a Python's Lunch, with $\chi_{\rm app}^{\rm MM}$ closer to the asymptotic boundary than $\chi_{\rm app}$ in a single state. This simply implies there are actually more exponentially complex operators.}$^{,}$\footnote{In this work we will often refer to the ``maximally mixed" state, $\rho$, of a given ensemble of pure states, $\lbrace \ket{\psi_i}: i \in I \rbrace$. By this, we just mean the \emph{averaged} state, $\rho = \frac{1}{|I|} \sum_i \ket{\psi_i}\bra{\psi_i}$. Depending on the ensemble, this density matrix, $\rho$, may or may not be proportional to the identity (i.e., maximally mixed in the standard sense).}

Let us briefly outline the intuition behind this essential insight. An immediate upshot of the Python's Lunch in the context of distinguishing complexity is that simple CFT operators cannot tell apart two bulk geometries that are identical at the level of effective field theory in the simple wedge but differ at leading order in $1/G_{N}$ inside the Lunch. In particular, such operators cannot discern if the appetizer surface is in fact the globally minimal QES that computes $S_{\rm vN}$ or just a local minimum. To identify whether a geometrically entangled state has an entropy of $\mathcal{O}(1/G_N)$, which would  distinguish between these two cases, we would have to check if the minimal QES is empty. If there were no Python's Lunch, this would be a computationally simple task --- but entanglement spoofers should be indistinguishable from highly entangled states. Thus, arguing by contradiction, at least one of the states in an entanglement spoofing pair must have a Python's Lunch. 

To illustrate this phenomenon concretely in AdS/CFT, we give an explicit construction of an ensemble of states that is likely pseudoentangled; it can be explicitly verified that this ensemble has a Python's Lunch in the maximally mixed state. The ensemble is a variant of the ensemble constructed by \cite{BouFef19}, which was argued to be pseudorandom and also has a Python's Lunch in the maximally mixed state~\cite{EngPen21b}.

Since a Python's Lunch must lie behind an event horizon \cite{EngFol24}, we find that black holes are the exclusive source of entanglement spoofing in geometrically entangled states in AdS. In a recent article \cite{EngFol24}, we found via complementary techniques that black holes are also the exclusive source of gravitational \textit{pseudorandomness} in AdS quantum gravity. Informally, an ensemble of unitaries $\{U_k\}$ (or states $\{\ket{\psi}_k\}$) is pseudorandom if a draw from such an ensemble is indistinguishable from a Haar random unitary (or state) by any low-complexity quantum algorithm (we will define such algorithms in Sec.~\ref{sec:prelims} below). The label `gravitational' refers to an additional requirement that the unitaries be \textit{sufficiently} pseudorandom on our code subspace; in particular, in \cite{EngFol24} we require the unitary to be pseudorandom by an amount inversely proportional to $G_N$. Much like back holes can spoof entanglement, they can also spoof randomness. This is not surprising: the lesson is that the black hole event horizon can essentially be used as a smokescreen to hide various properties of the state. Intuitively this is clearly true, and as it turns out it is also true in a quantitative way, at least for randomness and entanglement. See also a related discussion in~\cite{AkeEng22}. 

The paper is structured as follows. We begin in the next subsection by setting up notations and conventions. In Sec.~\ref{sec:review} we review the main ingredients that we will need: in Secs.~\ref{sec:EFI-pairs} and \ref{sec:pseudoentangled-states} we review EFI pairs and PES ensembles, respectively, and in Sec.~\ref{sec:geometric-states} we review some key ingredients from the framework of asymptotically isometric codes \cite{FauLi22}, which we use to formalize what we mean by geometric boundary states. In Sec.~\ref{sec:geometric-PE} we present our main results, the geometrization of entanglement-spoofing states. In Sec.~\ref{sec:comp-indist-outer-wedge} we argue that two geometric boundary states that are computationally indistinguishable must have identical outer wedges. In particular, this holds for geometric large-$N$ EFI pairs. In Sec.~\ref{sec:EFI-implies-PL}, we argue that at least one of the states of a large-$N$ EFI pair must have a Python's Lunch. In Sec.~\ref{sec:PES-implies-PL} we argue that geometric PES ensembles must have a Python's Lunch in the maximally mixed state over one of the ensembles; subsequently, in Sec.~\ref{sec:PES-construction}, we argue that geometric PES ensembles likely exist by providing an explicit candidate construction. In Sec.~\ref{sec:discussion}, we end with a short discussion of the main insights, our aspirations, and a few possible future directions. In App.~\ref{app:onenorm}, we provide details on a technical result that we need in Sec.~\ref{sec:EFI-implies-PL}.

\subsection{Preliminaries } \label{sec:prelims} 

\paragraph{Assumptions.} Throughout this paper, we will make the following assumptions. First and foremost, we will assume the strong Python's Lunch proposal \cite{BroGha19, EngPen21a,EngPen21b, EngWal17b, EngWal18}  explained above. Furthermore, we will assume that the spacetimes under consideration are AdS hyperbolic. We will also make use of the following genericity assumption: that the generalized entropies of different QESs of a complete boundary differ at leading order.

For concreteness, we assume that there is a parameter $N$ which controls the size of the boundary Hilbert space. We also assume there is a parameter $G_N$ scaling to zero in some way with $N$ (for concreteness we use $G_N \sim 1/N^2$, but our results hold for other scalings \textit{mutatis mutandis}), such that holographic states with non-trivial RT surfaces have entropies scaling like $A/G_N$ for $A$ some $\mathcal{O}(1)$ constant as $N \to \infty$. In our discussion, both $G_N$ and $A$ will be dimensionless, i.e., measured in units of the AdS length. We assume that the algebra of CFT operators at strictly infinite $N$ can be identified with the algebra of QFT operators on the spacetime emergent at infinite $N$ (see \cite{Har16, LeuLiu21a, LeuLiu22, FauLi22, PenWit23, Wit18} for a discussion of these algebras). We will be more explicit about our assumptions on the large-$N$ limit in Sec.~\ref{sec:geometric-states}.

\paragraph{Notation and conventions.} We now proceed to introduce and review some language and notation from the quantum computing literature. We will use $\mathcal{H}$ to denote Hilbert spaces and for a pure state $\ket{\psi}$, we will use $\psi$ to denote the density matrix $\ket{\psi}\bra{\psi}$. When we write $x \gets \mathcal{X}$, we mean a variable $x$ is drawn from a distribution $\mathcal{X}$. We use $\underset{x \gets \mathcal{X}}{\mathrm{Pr}}$ to denote taking the probability of some outcome occurring for $x$ drawn from $\mathcal{X}$. A function $f(x)$ is said to be \emph{negligible} in its parameter $x$ if for any constant $c>0$, $|f(x)| \leq x^{-c}$ for any sufficiently large $x$. A strict lower bound is indicated by $\omega (g(x))$, i.e., a function $f(x) \sim \omega (g(x))$ if for any $c>0$ there exist some $x_0 >0 $, such that $f(x) > c\, g(x)$ for all $x>x_0$. Similarly, we use $o (g(x))$ as a strict upper bound, i.e., a function $f(x) \sim o (g(x))$ if for any $c>0$ there exist some $x_0 >0 $, such that $f(x) < c \,g(x)$ for all $x>x_0$. The notation $f(x) \sim \Theta(g(x))$ denotes an upper and lower bound: there exist $c_1, c_2,x_0 >0$ such that $c_1 g(x) \leq f(x) \leq c_2 g(x)$ for all $x>x_0$. As usual in physics, $g \sim \mathcal{O}(f(x))$ means a non-strict upper bound: there exists some $c, x_0>0$ so that $g(x) \leq c f(x)$ for all $x \geq x_0$. Finally $\Omega(g(x))$ denotes a lower bound: $f(x)\sim \Omega(g(x))$ if there exists $c, x_0 >0$ such that $f(x) \geq c g(x)$ for all $x>x_0$.

The definitions of EFI pairs and PES ensembles make use of two concepts that we wish to clarify here: efficiently preparable and computationally indistinguishable. To understand their meaning, we briefly discuss quantum algorithms. A \textit{quantum algorithm} ${\cal A}$ is a map from an input Hilbert space ${\cal H}_{\rm in}$ to some output set ${\cal C}$. Note that the set ${\cal C}$ can be a Hilbert space, a classical space, or some other set altogether. For example, a quantum algorithm could take as input a quantum state $\ket{\psi}\in {\cal H}_{\rm in}$ and output a classical bit representation of an operator ${\cal O}$ (e.g., the coefficients of the representation of ${\cal O}$ in some fixed choice of basis). A quantum algorithm ${\cal A}$ acting on some $ {\cal H}_{\rm in}$ is said to be \textit{efficient} or equivalently \textit{computationally bounded} if it can be implemented by a quantum circuit of size no greater than polynomial in $\log \dim {\cal H}_{\rm in}$.\footnote{This involves some canonical choice of gates; however, this choice does not change the scaling in $\log \dim {\cal H}_{\rm in}$.} Here we will take $\mathcal{H}_{\rm in}$ to be the CFT Hilbert space, regularized by restricting to a sufficiently large microcanonical window. 

We say that a pair or ensemble of states is \textit{efficiently preparable} if there exists an efficient algorithm that can generate each state in the pair or ensemble.

Consider now an efficient algorithm designed to take in a state $\ket{\psi}$ and output $0$ if it determines that $\ket{\psi}$ was sampled from distribution $\mu_1$ and $1$ if it determines $\ket{\psi}$ came from $\mu_2$. We say that two such ensembles are \textit{computationally indistinguishable} if any efficient algorithm succeeds at this task with probability upper bounded by $1/2$ up to negligible corrections (i.e., exponentially suppressed corrections). An equivalent formulation is that the quantum algorithm outputs $1$ (or equivalently $0$) with almost the same probability when it is given a state from $\mu_{1}$, or from $\mu_{2}$. The distinguishing algorithms that are typically considered in the computer science literature are allowed ``advice" --- additional information that they can use to try to distinguish. For example, the distinguisher could be handed the expectation values of some operator in the two distributions that are to be distinguished.\footnote{It may seem like the ``advice'' could just provide the algorithm the answer to its distinguishing task. This is not the case: the advice is some arbitrary but \emph{fixed} information; it cannot depend on whether the given state $\ket{\psi}$ is from $\mu_1$ or from $\mu_2$, but it can, e.g., be information such as various statistics about the two distributions $\mu_1$ and $\mu_2$ under consideration.}

The definition of PES ensembles \cite{AarBou22} demands that such ensembles are indistinguishable (from highly-entangled state ensembles) even if the algorithm is allowed a polynomial number of copies of the state drawn from the ensemble. Having multiple copies of the state allows the algorithm to probe coherent information about the quantum state. EFI pairs, however, are technically only required to be indistinguishable for a single copy of the state. As we explain in Sec.~\ref{sec:review}, in this paper we will be interested in multi-copy indistinguishability for both pseudoentangled states as well as for EFI pairs (with slightly different notions of indistinguishability for states versus ensembles), thus minimally modifying the latter definition.

Finally, we introduce some notation related to bulk algebras of operators. In the classical limit, we will associate to bulk subregions algebras of operators, which one should think of as algebras of (bounded functions of) low energy fields in the bulk description. For generic subregions, these algebras will be of Type III$_1$ and so will not admit renormalizable density matrices or traces \cite{Sor23}. We will take the perspective that states reduced to an algebra $\mathcal{N}$ can be thought of as positive, linear functionals from the algebra to the complex numbers, which preserve the unit. Namely, they are linear maps $\rho: \mathcal{N} \to \mathbb{C}$ such that $\rho(a^{\dagger} a) \geq 0$ for $a \in \mathcal{N}$ and $\rho(1) = 1$. We will also utilize the assumptions of \cite{FauLi22}. In particular, all the states under consideration will be normal, but these assumptions will not be explicitly used anywhere in our work. 

We will also (briefly) need the use of quantum channels, which are normal, completely positive maps between von Neumann algebras. In particular, quantum channels take states to states under the pullback: if there is a quantum channel $\alpha: \mathcal{M} \to \mathcal{N}$ then for $\rho$ a state on $\mathcal{N}$, $\rho \circ \alpha$ is a state on $\mathcal{M}$.

Finally, we will need a notion of distance between different states on a von Neumann algebra. We will use the notion of the one-norm distance, which is given by
\begin{equation}\label{eqn:tracedist}
        ||\rho - \sigma||_1 = \sup_{a \in \mathcal{N}} \frac{|\rho(a) - \sigma(a)|}{||a||}\,, 
\end{equation}
where the supremum is over all self-adjoint operators in the algebra.

\section{Review of EFIs and Pseudoentangled States} \label{sec:review} 

In this section, we start by reviewing EFI pairs and pseudoentangled states in Secs.~\ref{sec:EFI-pairs} and \ref{sec:pseudoentangled-states}, respectively. These two notions were introduced in the quantum cryptography literature by~\cite{BraCan22} and \cite{AarBou22}, respectively; we here discuss them less formally, at a manner geared towards a physics audience. We will also define a strengthened version of EFI pairs, which will be more appropriate in the context of holography. In Sec.~\ref{sec:geometric-states} we formalize what we mean by geometric boundary states.

\subsection{EFI pairs}\label{sec:EFI-pairs}

We begin with EFIs, a notion defined in~\cite{BraCan22}. A pair of 
mixed quantum states is said to be an `EFI pair' if the states are (1) efficiently preparable (E); (2) statistically far (F); and (3) computationally indistinguishable (I) given a single copy. We will mostly consider a strengthened notion of EFIs, but we first give an informal description of standard EFIs (see~\cite{BraCan22} for a formal definition). 
EFI pairs $(\rho_\lambda, \sigma_\lambda)$ are indexed by a parameter $\lambda$ that is tied to the size of the Hilbert space on which the states are defined, i.e., $\log |\mathcal{H}| = \poly (\lambda)$, where $|\mathcal{H}|$ is the dimension of the Hilbert space $\mathcal{H}$. It will be convenient and sensible for us to use the CFT parameter $N$ to index these states, and we will do so in the remainder of this paper. 

The first requirement, (1) efficiently preparable, demands that $\rho_N$ and $\sigma_N$ can be prepared by some quantum algorithm of polynomial complexity (specifically, of $\poly(N)$ complexity), in the sense introduced in Sec.~\ref{sec:prelims}. Requirement (2) of being statistically far means that for any $N$, the trace distance between the two states $\rho_N$ and $\sigma_N$ must satisfy 
    \begin{equation}\label{eq:trpolybnd}
        ||\rho_N - \sigma_N||_1 \geq \frac{1}{{\rm poly} (N)}\,.
    \end{equation}
Requirement (3) of computational indistinguishability was described in Sec.~\ref{sec:prelims}. Here, (3) only requires such indistinguishability when the algorithm is given only a single copy of the state.

In this paper, we will consider a stronger notion of EFIs, which we will call large-$N$ multi-copy EFI pairs, or just large-$N$ EFI for short. This requires a stronger notion of the (2) statistically far and (3) computational indistinguishability properties. Let us begin with the strengthening of (3).    
Although there are good reasons to consider only single-copy indistinguishability in physical settings --- we often only have access to a single copy of the physical state of the system, e.g., observers jumping into black holes can do so only once --- we will find multi-copy indistinguishability a more useful notion to work with. Notably, we will see in Sec.~\ref{sec:comp-indist-outer-wedge} that multi-copy indistinguishability of two holographic states is sufficient to prove that the outer wedges of the states are identical. 

In addition to considering computational indistinguishability in the multi-copy sense, we will consider a version of EFIs where \eqref{eq:trpolybnd} is strengthened: specifically, we require the trace distance between $\rho_N$ and $\sigma_N$ to be $\Omega(1)$, instead of at least $\frac{1}{\poly(N)}$. Note that such EFI pairs are stronger `spoofers' --- they differ in trace distance by an amount that remains nonzero as $N\to\infty$ yet they still are indistinguishable to any polynomially complex quantum algorithm --- even with access to an $\mathcal{O}(1)$ number of copies. It may not be a priori obvious that such strong spoofers should exist; however, as we will discuss, we find it likely that these arise naturally in holography. We will call such pairs of states a \emph{large-$N$ multi-copy EFI pair} as defined below. 

\begin{defn}[Large-$N$ multi-copy EFI pair]\label{def:large-N-EFIs}  
Let $\rho_N$ and $\sigma_N$ be states in a Hilbert space $\mathcal{H}$ where $\log\dim\mathcal{H} = \poly(N)$. We say $(\rho_N, \sigma_N)$ is a large-$N$ EFI pair if the following criteria are satisfied:
    \begin{enumerate}
    \item Efficient preparation: There exists a quantum algorithm of $\poly(N)$ complexity that can produce the mixed state $\rho_N$, and similarly for $\sigma_N$.
    \item Statistically far: \begin{equation}
        ||\rho_N - \sigma_N||_1 \geq \Omega(1)\,. 
    \end{equation} 
    \item Computational indistinguishability: $\rho_N$ is computationally indistinguishable from $\sigma_N$ for any algorithm with advice and for any $\mathcal{O}(1)$ number of copies. 
    \end{enumerate}  
\end{defn} 

Note that it is important that the states $\rho_N$ and $\sigma_N$ are mixed: if they were pure, one could use the swap test to distinguish them. The swap test is a procedure that measures how much two quantum states differ. It takes as input two states $\psi$ and $\varphi$ and outputs a Bernoulli random variable that is 1 with probability $\frac{1}{2} + \frac{1}{2}\tr( \psi \varphi)$. 
Given even one copy of the unknown state $\chi \gets \{\rho_N,\sigma_N\}$, and given a copy of $\rho_N$ as advice --- which is generally permitted in the definition of EFIs --- a distinguisher could perform the swap test on $(\rho_N,\chi_N)$, which for pure states would always give $1$ if $\chi = \rho_N$ (but the probability would be lower for $\chi = \sigma_N$). So the states cannot be pure. 

It is then a reasonable question to ask \textit{how mixed} the states need to be. For the states to be indistinguishable, we need the swap test to output 1 for $\chi = \rho_N$ and $\chi = \sigma_N$ with the same probability; in other words, we need $\tr(\rho_N \sigma_N), \tr(\rho_N^2)$ and $\tr(\sigma_N^2)$ to be the same up to $\negl(N)$ corrections. We can then use the following inequality relating the one- and two-norms:  
    \begin{equation}
        || \rho_N - \sigma_N ||_1 \leq 2\sqrt{ \min({\rm rank}(\rho_N),{\rm rank}(\sigma_N))} \,||\rho_N - \sigma_N ||_2 \,.
    \end{equation}
The left hand side is lower bounded by the statistically far property of EFIs; the right hand side is upper bounded by the above arguments since 
    \begin{equation}
        || \rho_N - \sigma_N ||_2 = \sqrt{\tr (\rho_N - \sigma_N)^2 }= \sqrt{\tr (\rho_N^2) + \tr(\sigma_N^2) - 2 \tr(\rho_N \sigma_N)} = \negl (N)\,.
    \end{equation}
Thus we find 
    \begin{equation} \label{eq:EFIs-rank-requirement}
        \min({\rm rank}(\rho_N),{\rm rank}(\sigma_N)) \geq \frac{1}{\negl(N)}\,.
    \end{equation}
We will not need this bound for the main part of this paper; however, we will return to it in Sec.~\ref{sec:discussion}. 

\subsection{Pseudoentangled state ensembles}\label{sec:pseudoentangled-states}
We now move on to reviewing the second cryptographic notion: pseudoentangled state (PES) ensembles~\cite{AarBou22}.  
Informally, an ensemble of quantum states is pseudoentangled if the states are (1) efficiently preparable, and (2) indistinguishable from a second ensemble of states with much higher entanglement to any polynomially complex quantum algorithm, even if the algorithm is given polynomially many copies of the state. 
The intuitive goal of the definition is to say that a random draw from the first `pseudoentangled' ensemble is computationally indistinguishable from sampling the second ensemble of states. The essence of the pseudoentanglement notion relies on the idea that entanglement in general states cannot be ``felt'' by computationally bounded processes. 

Does this idea --- that entanglement is intangible in general states --- actually translate quantitatively into large possible differences between the entanglement of two indistinguishable ensembles? Suppose we have an ensemble $\lbrace \ket{\phi_k} \rbrace_{k}$, where $\ket{\phi_k} \in \mathcal{H}_A \otimes \mathcal{H}_B$, where  $ \mathcal{H}_A$ and $ \mathcal{H}_B$ are of comparable dimension. For example, if these are Hilbert spaces of qubit systems, then $| \mathcal{H}_A |\sim 2^{n}\sim | \mathcal{H}_B |$, i.e., they have approximately the same number of qubits $n$. In \cite{AarBou22} it was proven that it is possible to have states on $ \mathcal{H}_A \otimes  \mathcal{H}_B$ which are computationally indistinguishable but have parametrically large entanglement gap across the $A-B$ bipartition. 

However, analogously to the requirement that EFI pairs are mixed states, the states in the low entanglement ensemble cannot be a tensor product of two pure states in $\Hil_A$ and $\Hil_B$: access to multiple copies of the state allows one to use the swap test described in Sec.~\ref{sec:EFI-pairs}, which can easily tell a tensor product of pure states from a state with entanglement across $A-B$. It is straightforward to show that for two ensembles to be indistinguishable against efficient operations that use (polynomially) many copies of the state --- e.g., the swap test --- the von Neumann entropy across the bipartition must be bounded from below: it must grow with $| \mathcal{H}_A |$ as $\omega(\log \log | \mathcal{H}_A |)$, where we remind the reader that $\omega$ means a strict asymptotic lower bound. 
Again, this is analogous to the requirement on the ranks of EFI pairs \eqref{eq:EFIs-rank-requirement}. Ref.~\cite{AarBou22} explicitly constructed such ensembles: highly entangled states (across any bipartition) in one ensemble and low-entanglement states  getting arbitrarily close to the $\omega(\log \log | \mathcal{H}_A |)$ bound. The low-entanglement states were efficiently preparable and indistinguishable from the states in the former ensemble. Such states have been termed ``pseudoentangled''. We now present a (slightly informal) version of the definition in~\cite{AarBou22} here for pedagogical clarity. 

\begin{defn}[Pseudoentangled state ensembles]\label{defn:pes} 
Let $\mathcal{H} = \mathcal{H}_A \otimes \mathcal{H}_B$ be a bipartite Hilbert space, where $\log| \mathcal{H}_B|=\poly(\log |\mathcal{H}_A|)$. We say that the ensembles $\lbrace \ket{\psi_k} \rbrace_{k}$ and $\lbrace \ket{\phi_j} \rbrace_{j}$, where the indices $k$ and $j$ range over index sets of size $\exp({ \rm poly}(\log |{\cal H}_{A}|))$,  are \emph{pseudoentangled with gap $h(\log |{\cal H}_{A}|)$} if the following three conditions hold: 
\begin{itemize}
\item For any given $j, k$, the states $\ket{\psi_k}$ and $\ket{\phi_j}$ are efficiently preparable. 
\item With probability at least $1-\frac{1}{{ \rm poly}(\log |{\cal H}_{A}|)}$ over the choice of $\ket{\psi_k}$ and $\ket{\phi_j}$, the entanglement difference between the states reduced on $A$ (or equivalently $B$),  $|S(\psi_{k, A})-S(\phi_{j, A})|$, is $\Theta(h(\log |{\cal H}_{A}|))$.
\item The two ensembles are computationally indistinguishable.
\end{itemize}
We will call the ensemble with higher entanglement, $\{\ket{\psi_k}\}_k$, the entangled ensemble and the ensemble with lower entanglement, $\{\ket{\phi_j}\}_j$, the pseudoentangled ensemble.
\end{defn}
Note that since the entropy is upper bounded by the minimum of $\log|\mathcal{H}_A|$ and $\log|\mathcal{H}_B|$, PES ensembles can exist only when the minimum of $\log|\mathcal{H}_A|$ and $\log|\mathcal{H}_B|$ is at least the order of the gap function $h(\log |{\cal H}_{A}|)$.

As constructed by~\cite{AarBou22}, there are ``high gap" ensembles where the highly entangled ensemble has entanglement scaling with $\log |{\cal H}_{A}|$ and the pseudoentangled ensemble has entanglement (arbitrarily close to) $\log \log|{\cal H}_{A}|$. The pseudoentangled states in this construction were also pseudorandom states --- states that are computationally indistinguishable from Haar random. In general, pseudorandomness is not expected to be necessary to obtain pseudoentanglement, and the definition does not require it. In line with this, in this work we will present general results on holographic pseudoentanglement that do not require pseudorandomness; we also provide concrete candidates of high gap pseudoentangled states in holography, which \textit{are} pseudorandom.

\subsection{Geometric states }\label{sec:geometric-states}

As we just reviewed, the notions of EFI pairs and PES ensembles both require sequences of states in increasingly larger but finite-dimensional Hilbert spaces, labeled by some parameter $N$. To connect with results in AdS/CFT, we need to be precise about what we mean by boundary states whose bulk dual is geometric, since notions of geometry are really only defined asymptotically in the limit of large $N$. Thankfully, this is precisely the situation that the framework of \cite{FauLi22} was designed to handle. 

Accordingly, we will assume in this paper that our sequences of states will be described by the structure of an \emph{asymptotically isometric code}. Since we will not need the full power of the results of \cite{FauLi22}, we will use a slightly more compact notation to make this work more self-contained. Accordingly, we now highlight the elements of \cite{FauLi22} that we will make use of. These elements are incorporated into a list of assumptions that we will take to be part of the definition of a geometric state. 
\begin{assump}\label{ass:codespace}
    The bulk Hilbert space is given by a direct sum $\mathcal{H} = \mathcal{H}_{g_1} \oplus \mathcal{H}_{g_2} \oplus ...$, where $\mathcal{H}_{g_i}$ is the Hilbert space of QFT states defined perturbatively about a fixed background manifold $M_{g_i}$. Furthermore, the algebra of bounded operators on this Hilbert space breaks up into a direct sum over bounded operators on each Hilbert space individually, $\mathcal{B}(\mathcal{H}) = \bigoplus_i \mathcal{B}(\mathcal{H}_{g_i})$.
\end{assump}
Since we will mostly be concerned with distinguishing between two states, we will only need codes where $\mathcal{H}$ is the direct sum of at most two different $\mathcal{H}_{g_i}$'s, but we leave the notation general for now. Note that for us the geometries behave as separate superselection sectors. 
\begin{assump}\label{ass:isometry} 
There exists a sequence of bounded linear maps $V_N: \mathcal{H} \to \mathcal{H}_N$ such that for all $\ket{\psi}, \ket{\phi} \in \mathcal{H}$
\begin{align}
    \lim_{N \to \infty} \braket{\psi|V_N^{\dagger} V_N|\phi} = \braket{\psi|\phi}\,,
\end{align}
where $\mathcal{H}_N$ is the full boundary Hilbert space. We will often consider a situation where this Hilbert space breaks up into two copies of the Hilbert space of a single holographic quantum theory, $\mathcal{H}_N = \mathcal{H}_N^L \otimes \mathcal{H}_N^R$. We will assume that there is an entanglement wedge dual to $\mathcal{B}(\mathcal{H}_N^{L,R})$, with von Neumann algebra $\mathcal{N}^{L,R}$ which is reconstructible from $L,R$ in the sense of Lemma 8 of \cite{FauLi22}. From the results of \cite{FauLi22}, we will just need that for every state $\rho$ on $\mathcal{N}$ and every operator $a \in \mathcal{N}$, there is a sequence of states $\rho_N$ and boundary operators $a_N \in \mathcal{B}(\mathcal{H}_N)$ such that
\begin{align}
    & \lim_{N \to \infty} \rho_N(a_N) = \rho(a),
\end{align}
and where $\rho_N$ and $a_N$ are related by quantum channels to their bulk duals. Namely, there are channels $\alpha_N : \mathcal{B}(\mathcal{H}_N) \to \mathcal{N}$ and $\beta_N: \mathcal{N} \to \mathcal{B}(\mathcal{H}_N)$ such that $\rho_N = \rho \circ \alpha_N$ and $a_N = \beta_N(a)$. We assume all of these conditions hold for both $\mathcal{N}^L$ and $\mathcal{N}^R$.
\end{assump}
Since we will only be interested in a single entanglement wedge at a given time, we will usually drop the $L, R$ labels. Furthermore, we will always denote boundary operators and states with a subscript $N$; the corresponding bulk operators and states will lack the subscript. In this work, when we say that ``a CFT state $\rho_N \in \mathcal{B}(\mathcal{H}_N)$ is asymptotically dual to a bulk geometry," we will have in mind that $\rho_N = \rho \circ \alpha_N$ for some bulk state $\rho$ on $\mathcal{N}$. For brevity, we will often just say that such a sequence, $\lbrace \rho_N \rbrace_N$, is \emph{geometric}.

In what follows, we will be concerned with whether or not bulk observables lie in the outer wedge. We will denote the algebra associated to the outer wedge as $\mathcal{N}_{\mathrm{ow}}$. This algebra is a sub-algebra of $\mathcal{N} \supseteq \mathcal{N}_{\mathrm{ow}}$. Whether this inclusion is strict or not is the essential point of this work. If $\mathcal{N} \supsetneq \mathcal{N}_{\mathrm{ow}}$, then there exists a Python's Lunch. 

We now formalize assumptions about the complexity of the boundary dual of certain bulk operators.
\begin{assump}\label{ass:outwedge}
For all $a \in \mathcal{N}_{\mathrm{ow}}$, the sequence of dual operators $a_N$ have complexity scaling polynomially in $N$. Conversely, if for some $a \in \mathcal{N}$ the dual sequence $a_N$ is exponentially complex, then $a \notin \mathcal{N}_{\mathrm{ow}}$ and $\mathcal{N} \supsetneq \mathcal{N}_{\mathrm{ow}}$, so a Python's Lunch exists in at least one of the geometries $g_i$ in the decomposition $\mathcal{H}= \bigoplus_i \mathcal{H}_{g_i}$.
\end{assump}

We will need one final assumption on the existence of an area operator which measures entropies of boundary quantum states. 
\begin{assump}\label{ass:outerwedgearea}
    We will assume that there exists an area operator $\hat{A}^{\mathrm{RT}} \in \mathcal{N}$, which implies that there also exists an area operator $\hat{A}^{\mathrm{ow}} \in \mathcal{N}_{\mathrm{ow}}$ for the extremal surfaces that bound the entanglement wedge or outer wedge, respectively.\footnote{This follows because by construction, the algebra of the outer wedge ${\cal N}_{\rm ow}$ is also the algebra corresponding to the entanglement wedge of the coarse-grained ``simple'' state~\cite{EngWal17b, EngWal18, EngLiu23}. The assumption that the HRT surface area operator exists and lives in the algebra of the entanglement wedge, applied to the simple state, yields the desired result.} We will assume that the boundary dual $\hat{A}_N^{\mathrm{ow}} = \beta_N(\hat{A}^{\rm ow})$ of the outer wedge area operator is simple to measure from the boundary.\footnote{We expect that the simplicity of the area operator should follow from reconstruction near the boundary of the outer wedge \cite{EngWal17b,EngWal18,BroGha19, EngPen21a,EngPen21b}.} Furthermore, we assume that the HRT surface area measures the entropy of geometric boundary states asymptotically, i.e.,  
    \begin{align}
        \lim_{N \to \infty} S(\rho_N)4G_N = \lim_{N\to \infty}\rho_N(\hat{A}_N^{\mathrm{RT}}) = \rho(\hat{A}^{\mathrm{RT}})\,.
    \end{align}
\end{assump}

As for the application of these ideas to EFIs and PES ensembles, we will call a sequence of boundary states $\rho_N, \sigma_N$ that form a large-$N$ EFI pair as defined in Sec.~\ref{sec:EFI-pairs} a \emph{geometric EFI pair} if both $\rho_N$ and $\sigma_N$ are geometric. Furthermore, we call a sequence of ensembles $\lbrace \ket{\psi_{N,i}}\rbrace_i$ and $\lbrace \ket{\phi_{N,i}}\rbrace_i$, which form a sequence of PES ensembles, \emph{geometric pseudoentangled ensembles} if the average states of the ensembles 
\begin{align}
    \rho_N \equiv \frac{1}{|I'_N|} \sum_i \ket{\psi_{N,i}}\bra{\psi_{N,i}}\,, \qquad \sigma_N \equiv \frac{1}{|I_N|} \sum_i \ket{\phi_{N,i}}\bra{\phi_{N,i}}
\end{align}
are geometric. Here $I'_N$ and $I_N$ are the index sets of the two ensembles. $|I_N|$ and $|I'_N|$ are assumed to scale as $\Theta(e^{r N^{\gamma}})$ for $0 < \gamma <2$, with potentially different constants $r>0$ for each ensemble.\footnote{As becomes clear in Sec.~\ref{sec:PEproof}, it is also formally possible to carry through our proofs for $\gamma=2$, but in this case there is an $\mathcal{O}(1)$ upper bound on the constant $r$ for the $\sigma_N$ ensemble that we cannot determine. Furthermore, it is not clear whether backreaction of quantum fields are under control when $\gamma=2$.} The CFT Hilbert space (which will be taken to be an appropriate microcanonical window) has dimension $|\mathcal{H}_N| = e^{cN^2}$ for some $c>0$.

Having laid out our key assumptions and defined what we mean by geometric states, geometric EFI pairs and geometric pseudoentangled state ensembles, we now turn to proving the main claims of this paper. 

\section{Spoofing Entanglement and Pythons' Lunches} \label{sec:geometric-PE} 

In this section we present our main result: that entanglement-spoofing states in holography imply the existence of a Python's Lunch in the bulk dual. We start in Sec.~\ref{sec:comp-indist-outer-wedge} by arguing that geometric states that are (multi-copy) computationally indistinguishable must have identical outer wedges. Next, in Sec.~\ref{sec:EFI-implies-PL}, we combine this with the large trace distance property of geometric large-$N$ EFI pairs to conclude that one of the states in such an EFI pair must have a Python's Lunch in its bulk dual. We then turn to geometric PES ensembles in Sec.~\ref{sec:PEproof}. We first argue in Sec.~\ref{sec:PES-implies-PL} that the maximally mixed state on the lowly-entangled ensemble must have a Python's Lunch: if it did not, measuring the area operator would allow us to distinguish the two ensembles. We then give evidence for the existence of geometric PES ensembles in Sec.~\ref{sec:PES-construction} by providing an explicit candidate construction. 

\subsection{Indistinguishable states have identical outer wedges} \label{sec:comp-indist-outer-wedge}
We now argue that boundary states dual to semiclassical geometries, which are computationally indistinguishable with respect to $k$ copies, as described in \ref{sec:prelims}, must have identical outer wedges. By identical outer wedges, we mean that the states restricted to describe just the outer wedge\footnote{See \cite{EngWal18}  for a boundary protocol to obtain these reduced states.} have zero trace-distance, 
$D\equiv||\rho\vert_{\mathcal{N}_{\mathrm{ow}}} - \sigma \vert_{\mathcal{N}_{\mathrm{ow}}}||_1 = 0$.  

Consider any sequence of states $(\rho_\HN,\sigma_\HN)$ in $\mathcal{B}(\mathcal{H}_N)$ indexed by $N$, which are asymptotically dual to classical spacetimes with states $\rho, \sigma$ on the entanglement wedge $\mathcal{N}$.

Suppose now that the trace distance between $\rho$ and $\sigma$ does \textit{not} vanish on the outer wedge. We will show the contrapositive --- that this implies the states cannot be multi-copy computationally indistinguishable. Then for all $\varepsilon >0$,  there is an operator $O_{\varepsilon} \in \mathcal{N}_{\mathrm{ow}}$ such that
\begin{align}
    |\rho(O_{\varepsilon}) - \sigma(O_{\varepsilon})| \geq ||O_{\varepsilon}|| \times (D-\varepsilon)\,
\end{align}
as follows from \eqref{eqn:tracedist}.
Denote the boundary dual of this operator at finite $N$ by $O_{N,\varepsilon}$. By Assumption~\ref{ass:isometry}, there exists a function $\varepsilon'(N)>0$ with $\lim_{N \to \infty} \varepsilon'(N) = 0$ such that 
\begin{equation}\label{eqn:contradiction}
\begin{aligned}
\delta_{N,\varepsilon} \equiv |\rho_N(O_{N,\varepsilon}) - \sigma_N(O_{N,\varepsilon})| &\geq |\rho(O_{\varepsilon}) - \sigma(O_{\varepsilon})|\left(1 - \varepsilon'(N)\right) \\
&\geq ||O_{\varepsilon}|| \times \left(D-\varepsilon\right)\left(1 -\varepsilon'(N)\right) \\
& \geq \max \lbrace \varsigma_{\rho}(O_{\varepsilon}), \varsigma_{\sigma}(O_{\varepsilon}) \rbrace \left(D-\varepsilon\right)\left(1-\varepsilon'(N)\right)\,,
\end{aligned}
\end{equation}
where $\varsigma_{\rho}^2(O) \equiv \rho(O^2) - \rho(O)^2$ is the variance of the operator in the state $\rho$. In the third line, we used that $||O|| \geq \varsigma_{\psi}(O)$ for any state $\psi$ on an algebra.

We will show that this means that the states cannot be (multi-copy) computationally indistinguishable. Consider a quantum algorithm that is given $k$ copies of an unknown quantum state $\chi_N$ in $\mathcal{B}(\mathcal{H}_N)$, which may be either $\rho_N$ or $\sigma_N$. Let the algorithm now measure the operator $O_N$ on $\chi_N$ repeatedly $k$ times and then take the average of the measurements, obtaining a measurement outcome $o_N(k)$. Since $O \in \mathcal{N}_{\mathrm{ow}}$ is by definition supported only in the outer wedge, then by Assumption~\ref{ass:outwedge}, $O_N$ is only of polynomial complexity in the CFT~\cite{EngWal17b, EngWal18, EngPen21a, EngPen21b} and hence this operator is simple to measure on the boundary. The same is true of the average over $k$ measurements, as long as $k$ is not superpolynomial in $N$. 

Now assume without loss of generality that $\rho_N(O_N) \geq \sigma_N(O_N)$. If the outcome of the measurement $o_N(k) \geq \frac{\rho_N(O_N) + \sigma_N(O_N)}{2}$, then the algorithm guesses that $\chi_N = \rho_N$.\footnote{Note that the algorithm is allowed to have the expectation values $\rho_N(O_N), \sigma_N(O_N)$ as ``advice''.} Otherwise it guesses that $\chi_N = \sigma_N$. Now we would like to know the probability that this algorithm correctly guesses whether the state, $\chi_N$, is $\rho_N$ or is $\sigma_N$. Associated to each of the two states, there is a probability distribution associated to $o_N$, which we will denote as $\rho^{\otimes k}_N(o_N)$ and $\sigma^{\otimes k}_N(o_N)$. By the central limit theorem, the variances of these distributions are $\varsigma_{\rho_{N}}^2/k$ and $\varsigma_{\sigma_N}^2/k$, respectively, where $\varsigma_{\rho_{N}}^2$ and $\varsigma_{\sigma_{N}}^2$ are the variances of $O_N$ in the states $\rho_N$ and $\sigma_N$. Applying the so-called Cantelli inequalities (see, for example, \cite{Sav61}) to these distributions, we find that the probability that the algorithm guesses \textit{incorrectly} when given $\chi_N = \rho_N$ is 
    \begin{equation}
    \begin{aligned}
        \underset{o_N \leftarrow \rho_N^{\otimes k}}{\Pr}\left[o_N \leq \frac{\rho_N(O_N) + \sigma_N(O_N)}{2}\right] &= \underset{o_N \leftarrow \rho_N^{\otimes k}}{\Pr}\left[o_N - \rho_N(O_N) \leq - \frac{\delta_{N,\varepsilon}}{2}\right] \\
        &\leq \frac{\varsigma_{\rho_N}^2/k}{\varsigma_{\rho_N}^2/k + \delta_{N,\varepsilon}^2/4}\,,
    \end{aligned}
    \end{equation}
and the probability that the algorithm guesses incorrectly when given $\chi_N = \sigma_N$ is 
    \begin{equation}
    \begin{aligned}
        \underset{o_N \leftarrow \sigma_N^{\otimes k}}{\Pr}\left[o_N \geq \frac{\rho_N(O_N)+ \sigma_N(O_N)}{2}\right] &= \underset{o_N \leftarrow \sigma_N^{\otimes k}}{\Pr}\left[o_N - \sigma_N(O_N) \geq \frac{\delta_{N,\varepsilon}}{2}\right] \\
        &\leq \frac{\varsigma_{\sigma_N}^2/k}{\varsigma_{\sigma_N}^2/k + \delta_{N,\varepsilon}^2/4}\,.
    \end{aligned}
    \end{equation}
Using that for large enough $N$, there is a positive $\varepsilon''(N) \to 0$ with $N \to \infty$ such that $\varsigma^2_{\rho_N}$ and $\varsigma^2_{\sigma_N}$ obey  
\begin{align}
    &\varsigma_{\rho_N}^2 \leq \varsigma_{\rho}^2 \left(1+\varepsilon''(N)\right),\ \ \varsigma_{\sigma_N}^2 \leq \varsigma_{\sigma}^2 \left(1+\varepsilon''(N)\right),
\end{align} 
we get that the total probability that the algorithm guesses incorrectly is 
    \begin{equation}
        \Pr[{\text{algorithm is wrong}}] \leq \frac{1}{1 + \frac{k \left(D-\varepsilon\right)^2\left(1-\varepsilon'(N)\right)^2}{4(1+\varepsilon''(N))}}  ,
    \end{equation}
where we used the bound on $\delta_{N,\varepsilon}$ in  \eqref{eqn:contradiction}. We then find that for $k$ in the range
\begin{align}\label{eqn:kbound}
    k \geq \frac{4(1+\varepsilon''(N))}{\left(D-\varepsilon\right)^2\left(1-\varepsilon'(N)\right)^2} \left(1+ \frac{1}{\poly(N)}\right),
\end{align}
we get 
\begin{equation}
        \Pr[{\text{algorithm is wrong}}] \leq \frac{1}{2} - \frac{1}{\poly(\HN)}\,,
    \end{equation} 
and so the algorithm can distinguish between $\chi_N = \rho_N$ and $\chi_N = \sigma_N$ with better than $\negl(N)$ probability. Note that everything in the prefactor of equation \eqref{eqn:kbound} limits to a constant (namely, $4/D^2$) as $N \to \infty$. Thus, we can successfully distinguish the states as long as we are allowed any number of copies of the state which is $\mathcal{O}(1)$ in the large $N$ limit. Such a number of copies is allowed by our assumption of computational indistinguishability. So we find that if the trace distance does not vanish in the outer wedge, the states are not multi-copy indistinguishable; by the contrapositive, we conclude that if the states are multi-copy indistinguishable,  $D= ||\rho\vert_{\mathcal{N}_{\mathrm{ow}}} - \sigma \vert_{\mathcal{N}_{\mathrm{ow}}}||_1 = 0$.

\subsection{Geometric EFI pairs have a Python's Lunch} \label{sec:EFI-implies-PL}

We now argue that if any two boundary states have a large trace distance and are computationally indistinguishable, then at least one of them should have a Python's Lunch in its semiclassical bulk dual. 

Consider now two sequences of boundary states $\lbrace \rho_N \rbrace$ and $\lbrace \sigma_N \rbrace$, asymptotically dual to the bulk states $\rho, \sigma$ and computationally indistinguishable as in Sec.~\ref{sec:comp-indist-outer-wedge}, and furthermore assume $\rho_N$ and $\sigma_N$ have a large trace distance
\begin{equation}
        ||\rho_N - \sigma_N||_1 \geq \Omega(1)\,,
\end{equation}
for all sufficiently large $N$. By an argument detailed in App.~\ref{app:onenorm}, we know that for a sequence of boundary states dual to bulk states then the assumptions above imply
\begin{align}\label{eq:bulktracebnd}
    \lim_{N \to \infty} || \rho_N - \sigma_N||_1 = || \rho - \sigma||_1 \geq \Omega(1)\,.
\end{align}
To prove this, one needs to use the fact that there is a quantum channel which represents $\rho, \sigma$ on the boundary, namely $\rho_N = \rho \circ \alpha_N$ and $\sigma_N = \sigma \circ \alpha_N$, as mentioned in Assumption~\ref{ass:isometry}. Now, using the definition of the trace distance, \eqref{eq:bulktracebnd} gives us a bulk distinguishing operator $a^*$ whose expectation value differs in $\rho, \sigma$ by some non-zero amount. Since the states are computationally indistinguishable, by the argument of Sec.~\ref{sec:comp-indist-outer-wedge}, it cannot be the case that $a^* \in \mathcal{N}_{\rm ow}$ or else $||\rho \vert_{\mathcal{N}_{\rm ow}} - \sigma\vert_{\mathcal{N}_{\rm ow}}||_1 \geq \frac{|\rho(a_*) - \sigma(a_*)|}{||a_*||} >0$. Therefore, $\mathcal{N}_{\rm ow} \subsetneq \mathcal{N}$ and we have a Python's Lunch in at least one of $\rho$ or $\sigma$.

\subsection{Geometric pseudoentangled states \& Pythons'  Lunches}\label{sec:PEproof}

We now discuss the holographic interpretation of PES ensembles. Concretely, we show that geometric PES ensembles with entanglement gap $h(\log|\Hil_A|) = \Theta(N^2)$ have a Python's Lunch in the maximally mixed state over the low-entanglement ensemble.

\subsubsection{Geometric pseudoentangled states imply Python's Lunch}\label{sec:PES-implies-PL}
Consider two ensembles of boundary states  $\lbrace \ket{\psi_{N,i}}\rbrace_{i\in {I'_N}}$ and $\lbrace \ket{\phi_{N,i}} \rbrace_{i\in {I_N}}$, which form a pair of PES ensembles. We can characterize these two ensembles in terms of their average density matrices: 
\begin{align}
    \tilde{\rho}_N \equiv \frac{1}{|I'_N|}\sum_{i\in I'_N} \ket{\psi_{N,i}}\bra{\psi_{N,i}}, \ \ \tilde{\sigma}_N \equiv \frac{1}{|I_N|}\sum_{i\in I_N} \ket{\phi_{N,i}}\bra{\phi_{N,i}}.
\end{align}
As in the previous subsections, we include a subscript $N$ to remind the reader that these ensembles depend on $N$. Recall from Sec.~\ref{sec:geometric-states} that $\log|I_{N,N'}| = \Theta(N^{\gamma})$ with $0 < \gamma < 2$.

We assume that these two ensembles have a large $N$ holographic dual, namely that there are states $\tilde{\rho}$ and $\tilde{\sigma}$ acting on the bulk such that
\begin{align}
\tilde{\rho}_N = V_N\tilde{\rho}V_N^{\dagger}\,, \  \ \tilde{\sigma}_N = V_N\tilde{\sigma}V_N^{\dagger}\,,
\end{align} 
where $V_N$ is the bounded linear map introduced in Assumption~\ref{ass:isometry}. Recall that the states $\ket{\psi_{N,i}}$, $\ket{\phi_{N,i}}$ are each bipartite, say over $\Hil_L\otimes \Hil_{R}$. Let
$$\rho_{N,i} \equiv \tr_R(\ket{\psi_{N,i}}\bra{\psi_{N,i}})\,,\ \ \sigma_{N,i} \equiv \tr_R(\ket{\phi_{N,i}}\bra{\phi_{N,i}})\,,$$
$$\rho_N \equiv \tr_R(\tilde{\rho}_N)\,,\ \ \sigma_N \equiv \tr_R(\tilde{\sigma}_N)\,.$$ 
In what follows, we will just consider states over $\Hil_L$.   
Now, by the assumption that the ensembles $\lbrace \ket{\psi_{N,i}}\rbrace$ and $\lbrace \ket{\phi_{N,i}} \rbrace$ are pseudoentangled, we know that $\rho_N$ and $\sigma_N$ have a large gap in their von Neumann entropies. They are also indistinguishable for a single copy, by the indistinguishability property of the pseudoentangled ensembles invoked for a single copy. Assume, without loss of generality, that the ensemble represented by $\sigma_N$ has lower entanglement. 

To proceed, we start from Assumption~\ref{ass:outerwedgearea}, i.e., we assume that we can measure the operator $\hat{A}^{\mathrm{ow}}\in {\cal N}_{\rm ow}$, which corresponds to the area of the extremal surface bounding the outer wedge, and that this area's boundary dual $\hat{A}_N^{\rm ow}$ is simple to measure (observables in the outer wedge were argued to be simple in~\cite{EngWal18, EngPen21a}). Let $\hat{A}^{\rm RT}$ be the operator corresponding to the area of the HRT surface (note that we do not need to invoke the complete QES prescription, since we are computing the area and working at leading order in $N$, under the assumption that bulk entropy is subleading). By the HRT prescription, we know that 
\begin{align}\label{eqn:RT}
    \rho(\hat{A}^{\mathrm{ow}}) \geq \rho(\hat{A}^{\mathrm{RT}}) \equiv A_{\rho}\,,
\end{align} 
and analogously for the $\sigma$ state.
Since the states $\rho$ and $\sigma$ are classical, we know that the geometry has zero fluctuations associated to the area operator; they lie in superselection sectors associated to eigenvalues of the area operator \cite{Har16, ChaPen22}. 

To proceed, assume for contradiction that $\sigma$ does not have a Python's Lunch; note that this implies $\sigma(\hat{A}^{\rm ow}) = \sigma(\hat{A}^{\rm RT})$. 
We then have 
        \begin{align} \nonumber
            \lim_{N\to \infty} |\rho_N(\hat{A}_N^{\rm ow}) &- \sigma_N(\hat{A}_N^{\rm ow})| = | \rho (\hat{A}^{\rm ow}) - \sigma(\hat{A}^{\rm ow}) | \\ \nonumber
            &\geq | \rho (\hat{A}^{\rm RT}) - \sigma(\hat{A}^{\rm RT})|\\
            &= \lim_{N\to\infty} 4G_{N}(S(\rho_N) - S(\sigma_N)) \\ \nonumber
            &\geq \lim_{N\to\infty} \Bigg[\frac{4G_N}{|I_N'|} \sum_{i\in I'_N}S(\rho_{N,i})- \frac{4G_N}{|I_N|}\sum_{i\in I_N} (S(\sigma_{N,i}) + \log|I_N|) \Bigg]\\ \nonumber
            &\geq \lim_{N\to\infty} \left[  \left( 1 - \frac{1}{\poly(N)} \right) \frac{h(\log |\Hil_A|)}{\Theta(N^2)} - \frac{1}{\poly(N)}\frac{\log|\Hil_A|}{\Theta(N^2)} - \frac{\log|I_N|}{\Theta(N^2)} \right] \\ \nonumber
            &\geq c\,,
        \end{align}
where in the last line, we used the entanglement gap $h(\log|\Hil_A|) = \Theta (N^2) \geq cN^2$ for some constant $c>0$, and that $|I_N|=e^{\Theta(N^\gamma)}$ with $\gamma < 2$. In the first line, we used the asymptotic code structure; in the second line, the assumption that $\sigma$ does not have a Python's Lunch, i.e., that $\rho(\hat{A}^{\rm ow}) \geq \rho(\hat{A}^{\rm RT})$ and $\sigma(\hat{A}^{\rm ow}) = \sigma(\hat{A}^{\rm RT})$; in the third, the HRT prescription and the assumption that $\rho_N$ has high, and $\sigma_N$ low entanglement to remove the absolute value; in the fourth, concavity of the entropy; and in the fifth, the assumption of pseudoentanglement. In particular, we used that pseudoentangled ensembles are such that $\geq 1-\frac{1}{\poly(\HN)}$ fraction of the states in the ensembles have the entropy gap $h(\log|\Hil_N|)$ and for the remaining $\leq \frac{1}{\poly(\HN)}$ fraction, we simply have the trivial bounds of $S(\rho_{N,i}) \geq 0$ and $S(\sigma_{N,i}) \leq \log|\Hil_A|$.    

By assumption, $\hat{A}_N^{\rm ow}$ is simple to measure on the boundary, and by Assumption~\ref{ass:isometry}, the fluctuations in $\hat{A}_N^{\rm ow}$ go to zero as $N \to \infty$ in both $\rho_N$ and $\sigma_N$, since they are zero in $\rho$ and $\sigma$. So we see that the states $\rho_N$ and $\sigma_N$ are then clearly distinguishable on the boundary, by measuring $\hat{A}_N^{\rm ow}$ for sufficiently large $N$. This contradicts the single-copy computational indistinguishability of $\rho_N$ and $\sigma_N$. Since we only assumed that $\sigma$, corresponding to the ensemble of low-entanglement states, has no Python's Lunch, we have thus shown that at least $\sigma$ contains a Python's Lunch. (In general, $\rho$ may or may not contain a Lunch.)

\subsubsection{A construction of geometric pseudoentangled states} \label{sec:PES-construction}
We have argued that if geometric pseudoentangled states exist, then a Python's Lunch must exist, in the sense that it exists in the maximally mixed state over the lowly entangled ensemble.  We now proceed to give evidence that geometric PES ensembles do in fact exist. Note that, strictly speaking,  the relevant QES in this example is not a classical HRT surface, and due to time evolution over times increasing with $N$, bulk geometric features might depend on $N$, meaning that the strict $N\rightarrow \infty$ limit is subtle. Previous sections applied only to classical extremal surfaces, but since the entropy term here is still subleading to the area term, we expect similar arguments to hold in this context. 

Consider a collection of $m$ different flavors of simple unitary operators $U_{1}, \ldots , U_m$ that each inject an $\mathcal{O}(1)$ amount of energy near the boundary of AdS. Assume furthermore that each $U_i$ factorizes and acts identically on the two CFTs --- the former ensuring that it does not alter the entanglement between the CFTs. Working in the Schr\"odinger picture, let $\ket{\chi_0}$ be some initial geometric seed state and construct an ensemble of $m^{\ell}$ states given by 
\begin{equation}
\begin{aligned}
    \ket{\chi_{\bm{i}}} = e^{-iHt} U_{i_{\ell}}e^{-iHt} U_{i_{\ell-1}} e^{-iHt}
    U_{i_{\ell-2}} \ldots U_{i_{3}} e^{-iHt}
    U_{i_{2}}
    e^{-iHt} U_{i_1} \ket{\chi_0}\label{eq:BFV}
\end{aligned}
\end{equation}
where each $i_{j}\in \{1,\ldots, m\}$ and the index $\bm{i}=(i_1,\ldots,i_\ell)$. Here $t$ is some fixed time interval larger than twice the scrambling time. These types of states were studied in the context of pseudorandom state ensembles by Bouland, Fefferman, and Vazirani (BFV) in \cite{BouFef19} (see also \cite{SheSta14, SheSta14b}) and were shown to feature a Python's Lunch in the maximally mixed state in \cite{EngPen21b}. The ensemble $\{\ket{\chi_{\bm{i}}}\}_{\bm{i}}$ was argued to be indistinguishable from a Haar random ensemble for any polynomial number of copies --- even when $\ell=o(N^2)$. However, this indistinguishability  relies on a computational model that does not allow the use of simple measurements to determine what $U_{i_{\ell}}$ was applied to the state $e^{iHt}\ket{\chi_{\bm{i}}}$. We will first assume this is the case and then discuss the physical interpretation of this assumption in AdS/CFT below. 

To construct our PES ensemble, we now pick two particular seed states $\ket{\phi_0}$ and $\ket{\psi_0}$ such that the two generated BFV-type ensembles \eqref{eq:BFV} are candidates for PES ensembles. We take $\ell =\Theta(N^{\alpha})$  for any $0<\alpha<2$. For $\ket{\psi_0}$ we simply take the thermofield double dual to a large black hole of mass/CFT energy $M=\Theta(N^2)$.\footnote{Projected to a microcanonical energy window if so desired.} For the state $\ket{\phi_0}$ we take two identical one-sided large AdS black holes formed from collapse and evolved for a time much larger than the scrambling time, so they have settled down completely. We also ensure that the two identical collapsing matter distributions have an amount of entanglement that is $\omega(\log N)$ and $o(N^2)$, so the two black holes will share this amount of entanglement. We pick our collapsing black holes to have the same mass as the thermofield double (to polynomial precision in $1/N$). We want to show $\Phi=\{\ket{\phi_{\bm{i}}}\}$ and $\Psi=\{\ket{\psi_{\bm{i}}}\}$ are PES ensembles. Clearly, the states in these ensembles are efficiently preparable and geometric.\footnote{Assuming the thermofield double is efficiently preparable, which is believed to be the case \cite{CotFre19, WuHsi19,SagPre21, Su21}.} Indistinguishability was argued by BFV, under the assumption that there is no simple way to measure what shock was applied to the state $e^{i H t} \ket{\chi_{\bm{i}}}$.  
BFV already argued that $\Psi$ is computationally indistinguishable from Haar random, and they sketched a proof in a toy qubit model of AdS/CFT. Their argument did not rely on the particular entanglement structure of the thermofield double state, so their arguments appear equally valid for the ensemble $\Phi$ as well, which only differs in the choice of initial seed state. By transitivity of indistinguishability, the two ensembles $\Phi$ and $\Psi$ would then be indistinguishable from each other, and hence pseudoentangled ensembles.\footnote{As pointed out in BFV, there may be subtleties, that do not seem to be an issue in our context, in whether their ensemble is indistinguishable \emph{from the Haar random ensemble}, which was the focus of their work. However, in our case, the initial seed states and thus states in the ensembles all have approximately the same energy, so we expect the two ensembles here may be indistinguishable \emph{from each other}, which suffices for pseudoentanglement.} 

Let us now discuss the physical interpretation of the assumption that the last shock cannot be measured. Taking for example $\ket{\chi_0}$ to be dual to a large black hole formed from collapse, the physical preparation of the states \eqref{eq:BFV} is as follows: each operator $U_{i_j}$ injects some $\mathcal{O}(1)$ amount of energy that falls into the black hole upon forwards time evolution. 
We always consider reflecting AdS boundary conditions, so observers residing near the AdS boundary in (the time evolution of) the state
\begin{equation}
\begin{aligned}
U_{i_j}e^{-iHt} U_{i_{j-1}}\ldots U_{i_2} e^{-iHt}U_{i_1}\ket{\chi_0}
\end{aligned}
\end{equation}
would see the following: as they approach the insertion time of $U_{i_j}$ from the past, an excitation travels outwards towards the boundary, bounces off it, and then falls into the black hole. Naively, we might think that the same happens for all the earlier excitations created by $U_{i_{j-1}},U_{i_{j-2}},\ldots,U_{i_1}$; however, as realized by \cite{SheSta14b}, this is not the case. As long as $t$ is greater than approximately the scrambling time of the black hole, all the other shocks will be hidden inside the black hole. 
However, the last shock is still visible near the boundary. If the $U_{i}$ created orthogonal states (when acting on this ensemble) labeled by $i$, for example realized by having $U_{i}$ insert local lumps of energy near the physical boundary at distinct physical locations, we could use a simple localized energy measurement to determine which $U_i$ was applied. In that case, we could apply $e^{iHt}U_i^{\dag}$ and then repeat the procedure, gradually unwinding the applied shocks with an algorithm of complexity $\mathcal{O}(m \ell)$. To get around this, we can assume that the different $U_i$ do not commute, so that simple measurements near the boundary cannot be used to determine with certainty which $U_i$ was applied. This means that we cannot have that $i$ labels physical location or flavor eigenstates. However, it could for example instead label different non-orthogonal superpositions of these. 

\section{Discussion} \label{sec:discussion}
In this paper, we have shown that two recently proposed notions of entanglement spoofers in the CFT constrain the geometry of the AdS bulk. Working to leading order in $1/N$ and assuming the strong Python's Lunch conjecture, we first showed that boundary states that are multi-copy computationally indistinguishable must have identical outer wedges. Using this, we argued that at least one of the states of a geometric large-$N$ EFI pair must have a Python's Lunch in its bulk dual. Next, we showed that if $\{\ket{\psi_i}\}_i, \{\ket{\phi_i}\}_i$ constitute a geometric PES ensemble with a large entanglement gap, then the low-entanglement ensemble must have a Python's Lunch in the maximally mixed state over the ensemble.  Finally, we pointed out that a simple modification of the BFV construction of pseudorandom states likely yields an explicit example of geometric pseudoentangled states. 

We now turn to a discussion of consequences, possible extensions, and subtleties in our results. 

\subsubsection*{A converse?}
It is natural to ask whether two states $(\rho, \sigma)$ that have classically different Pythons, but agree in their outer wedges, always constitute a large-$N$ EFI pair (and thus also a standard EFI pair). Naively, this appears to be true: the bulk states only differ inside the Python, which is exponentially hard to reconstruct. However, it cannot in fact be true without additional assumptions. As shown by the minimal rank condition \eqref{eq:EFIs-rank-requirement}, derived by imposing indistinguishability under the boundary swap test, EFIs must be mixed states, and this fact was not implied by the na\"ive argument just given. However, if we additionally impose the minimal rank condition, it could in principle be that the converse holds, although we do not have an argument at this stage.  

\subsubsection*{All-cuts version of pseudoentanglement}
As alluded to in the introduction, there exists a more general class of PES ensembles than the one considered in this paper. Concretely, in the second version of the work of \cite{AarBou22}, the definition of PES ensembles was expanded to require that the states are pseudoentangled across almost all partitions of the system. In this work, we followed the first version of \cite{AarBou22}, in which the states are pseudoentangled across a single, fixed cut. It might be tempting to also consider this stronger definition, but given the highly constrained structure of entanglement in finite energy CFT states and the resulting constrained asymptotic behavior of quantum extremal surfaces, there cannot exist a geometric bulk description of an all-cuts pseudoentangled holographic state with a large gap (i.e., scaling as $\log |{\cal H}_{A}|$ to any positive power). This is due to the fact that all finite energy CFT states look the same sufficiently deep in the UV. 
In the bulk, this means that once a region $R\subset A$ is sufficiently small, the QES resides far out in the asymptotic region, and in particular, completely in the causal wedge of $A$. An entropy difference scaling as $(\log |{\cal H}_{A}|)^\delta$ with $\delta >0$ means a difference in entropy that scales as a power of $G_N^{-1}$. This would have to manifest as a difference in the geometry of the EFT bulk state in the causal wedge that is within the perturbative regime. Since causal wedge reconstruction is expected to  be simple, this would break computational indistinguishability. It is in principle possible that there is some version of pseudoentanglement for some privileged cuts (but not all cuts); we remain agnostic on this point.  

\subsubsection*{Connection to black hole information}
It has been argued in \cite{Bra22} that the existence of EFI pairs is equivalent to the hardness of decoding Hawking radiation --- that is, if and only if EFI pairs exist, there exist ``hard-to-decode radiation states". These latter states are defined to be pure states $\ket{\psi}_{HBR}$ on a Hilbert space that factorizes in three registers: the black hole interior ($H$), the (old) radiation outside of the black hole ($R$), and the newly outgoing radiation ($B$). The claim is that the existence of EFI pairs is equivalent to the existence of a decoder that takes $R$ as input and outputs a qubit that is maximally entangled with $B$. This decoder is, however, inefficient: decoding the Hawking radiation is strongly believed \cite{HarHay13} to be exponentially difficult.\footnote{This has been known since \cite{HarHay13}; the new claim in \cite{Bra22} is the existential equivalence relation to EFI pairs.} This is true for single but not multiple copies of the Hawking radiation at very late times, with only $\omega (\log G_{N})$ entropy remaining in the black hole, due to the swap test. At intermediate times between the Page time and this very late time regime, it is reasonable to expect that the radiation is multi-copy indistinguishable: the radiation is sufficiently mixed that the swap test on its own may not be able to distinguish between the states. Indeed, it is suggestive that the post-Page time entanglement wedge of the radiation does, in fact, have a Python's lunch. The extent to which the intermediate-late regime Hawking radiation is well-modeled by our notion of multi-copy EFI pairs is an interesting open question that we leave to future work. 

\subsubsection*{Dustballs and states without a large-$N$ limit}
A natural way to attempt a counterexample to the result that large-$N$ EFIs imply Pythons is by putting EFIs directly into the bulk. However, these types of states are excluded by our assumptions. In particular, they do not appear to have a strict large--$N$ limit, and there is no well-defined classical background on which these states can be said to live. A concrete example would be the dustball spacetimes of \cite{AkeLei19, AkePen20}, which contain large dustballs hosting an $\mathcal{O}(N^2)$ number of internal states, but where the pointwise energy density is small enough so that the local backreaction is under control as $N\rightarrow \infty$. The dustball quickly collapses into a black hole, but, as shown in \cite{AkeLei19}, the parameters of the dustball can be tuned such that the horizon does not intersect the time-symmetric slice. This means that there can be no Python's Lunch in this spacetime, but since the entropy of the configuration scales with $N^2$ it seems reasonable that one could find an EFI pair associated to the internal states of the dust. This would essentially correspond to putting EFI pairs directly into the causal wedge. However, either these states do not have a well defined large-$N$ limit, or the large-$N$ limit is not described by the usual paradigm of a classical asymptotically AdS background with free or perturbatively interacting quantum fields on top.  In particular, the asymptotically isometric code picture adopted in this paper is not applicable. If $O_N$ was the boundary dual to the bulk operator distinguishing $\rho_N$ and $\sigma_N$ at finite $N$, then we would not expect to have an asymptotically isometric code where we can write $O_N = \beta_N(O)$ for some bulk operator $O$ that acts on $\mathcal{H}$. Additionally, the geometry has non-trivial features up to radii of order $r\sim \mathcal{O}(N^2)$, and in the limit $N\rightarrow \infty$ we do not get an asymptotically AdS geometry with the same AdS radius as the CFT vacuum. 

A similar thing can be said of an old evaporating black hole; in this case, the volume of a Cauchy slice is
$\mathcal{O}(N^2)$ due to the linear growth of the wormhole interior for a time of $\mathcal{O}(t_{\rm Page})$. This possibly precludes us from applying our results to the evaporating black hole context. Nevertheless, we expect that the evaporating black hole is fundamentally different from the dustball example. The reason is that the operator which distinguishes the EFI states in the dustball case likely needs to be exponentially complex \emph{in the bulk}. Such operators are not believed to be well-defined in the bulk EFT per arguments in \cite{AkeEng22}, and as remarked on above. By contrast, the operators behind the Python's Lunch of an evaporating black hole have exponential complexity in the boundary while being perfectly fine bulk EFT operators. Furthermore, the volume growth in the $N\rightarrow \infty$ limit takes place in the interior, and therefore does not break the AdS asymptotics. 

\subsubsection*{Perturbative $1/N$ corrections}
The asymptotically isometric codes framework \cite{FauLi22} adopted here provides an embedding of the strict $N\rightarrow \infty$ bulk Hilbert space into the finite-$N$ CFT Hilbert space. This means that in the bulk we have not incorporated perturbative in $1/N$ corrections beyond the leading part that describes QFT on a rigid background (including free gravitons). We do however suspect that a different version of our result that computational indistinguishability implies equality of outer wedges is true with perturbative corrections. Namely, we expect that if we require computational indistinguishability for any $\poly(N)$ number of copies, then the outer wedges must agree to all orders in $1/N$ perturbation theory. The argument is expected to be similar to the one given in Sec.~\ref{sec:comp-indist-outer-wedge}, except that our assumption leading to a contradiction would be that $||\rho|_{\mathcal{N}_{\rm ow}}-\sigma|_{\mathcal{N}_{\rm ow}}||_1 \geq \frac{1}{\poly(N)}$, giving that there always is a sufficiently large but polynomial number of copies that can break indistinguishability (given by \eqref{eqn:kbound} with $D=1/\poly(N)$). The idea is simply that with a $\poly(N)$ number of copies, through repeated measurements, we can reduce the variance of any observable to be polynomially small for any polynomial, giving polynomially accurate determination of all outer wedge expectation values. We also expect a Python's Lunch, since otherwise the outer wedge is everything, implying that $||\rho -\sigma||_1 = \text{negl}(N)$, contradicting statistical distinguishability (independently of whether or not we allow $||\rho -\sigma||_1=1/\text{poly}(N)$).

\appendix 

\section*{Acknowledgements}
It is a pleasure to thank Raphael Bousso, Daniel Harlow, Thomas Faulkner, and Arvin Shahbazi-Moghaddam for valuable discussions. This work is supported in part by the Department of Energy under Early Career Award DE-SC0021886 (NE, \AA{}F, and LY) and QuantISED DE-SC0020360 contract no. 578218 (NE), by the Heising-Simons Foundation under grant no. 2023-4430 (NE and AL), the Sloan Foundation (NE), the Templeton Foundation via the Black Hole Initiative (NE and EV), the Gordon and Betty Moore Foundation (EV), the Packard Foundation (AL), and the MIT department of physics (NE). The opinions expressed in this publication are those of the authors and do not necessarily reflect the views of the John Templeton or Moore Foundations.

\section{Asymptotic One-Norms}\label{app:onenorm}
In this Appendix, we prove that the one-norm difference between two sequences of holographic states limits to the one-norm difference between their bulk duals.\footnote{We thank Tom Faulkner for explaining this proof to us.}

Consider two sequences of boundary states $\rho_N$ and $\sigma_N$ on the boundary algebra $\mathcal{N}_N$. In the main text, we take $\mathcal{N}_N = \mathcal{B}(\mathcal{H}_N)$, but we will be more general here. Assume $\rho_N$ and $\sigma_N$ are asymptotically dual to bulk states $\rho, \sigma$ on the bulk algebra $\mathcal{N}$. By the discussion in Sec.~\ref{sec:geometric-states}, we know that these states are related to the bulk states via a quantum channel, $\alpha_N: \mathcal{N} \to \mathcal{N}_N$.

Furthermore, by Assumption~\ref{ass:isometry}, for any operator $a \in \mathcal{N}$, there is a quantum channel, $\beta_N$, which maps to a boundary operator $a_N \in \mathcal{N}_N$. Then by monotonicity of the one-norm under action of a channel, we have for any $a \in \mathcal{N}$
\begin{equation}
\begin{aligned}
    ||\rho_N - \sigma_N||_1 &= ||\rho \circ \alpha_N - \sigma \circ \alpha_N||_1 \nonumber \\
    &\geq || \rho \circ \alpha_N \circ \beta_N - \sigma \circ \alpha_N \circ \beta_N||_1 \\
    &\geq \frac{|\rho(\alpha_N(\beta_N(a))) - \sigma(\alpha_N(\beta_N(a)))|}{||a||}\,.
\end{aligned}
\end{equation}
Taking the lim sup of both sides and then optimizing over $a$, we arrive at 
\begin{align}
    \limsup_{N\to\infty} ||\rho_N - \sigma_N||_1 \geq  ||\rho - \sigma||_1\,.
\end{align}
We can also arrive at the same thing for the $\liminf$ of both sides so that 
\begin{align}
    \liminf_{N\to\infty} ||\rho_N - \sigma_N||_1 \geq  ||\rho - \sigma||_1\,.
\end{align}
On the other hand, by the monotonicity of the trace distance, we have that for any $N$,
\begin{align}
    ||\rho_N -\sigma_N||_1 = ||\rho \circ \alpha_N - \sigma \circ \alpha_N ||_1 \leq ||\rho - \sigma||_1\,.
\end{align}
Again taking the $\limsup$ and $\liminf$ of both sides, we see that
\begin{align}
    \lim_{N\to \infty} ||\rho_N - \sigma_N||_1 = ||\rho - \sigma||_1\,,
\end{align}
as desired. 

\bibliographystyle{jhep}
\bibliography{all}

\end{document}